\documentclass{article}
\usepackage{amsfonts}
\usepackage{amsmath}
\usepackage{amssymb}
\usepackage[dvips]{color,graphicx}
\usepackage{slashed}

\newcommand{\be}{\begin{equation}}
\newcommand{\ee}{\end{equation}}

\newcommand{\ba}{\begin{equation}\begin{array}}
\newcommand{\ea}{\end{array}\end{equation}}

\newcommand{\bea}{\begin{eqnarray}}
\newcommand{\eea}{\end{eqnarray}}

\newcommand{\bse}{\begin{subequations}}
\newcommand{\ese}{\end{subequations}}

\newcommand{\Too}{T^{\hat{0}\hat{0}}}
\newcommand{\Tab}{T^{\hat{a}\hat{b}}}

\newcommand{\rr}{r}
\newcommand{\rt}{r}

\newcommand{\rA}{\rr_{\scriptscriptstyle{A}}}
\newcommand{\rB}{\rr_{\scriptscriptstyle{B}}}

\newcommand{\ds}{\displaystyle}

\newcommand{\ZZ}{{\declareslashed{}{\text{-}}{0.02}{0}{Z}\slashed{Z}}}

\title{Bounded, asymptotically flat toroidal exteriors for cylindrically symmetric spacetimes}
\author{A J Janca\\\footnotesize{Department of Physics, North Carolina State University\\Raleigh NC 27695, United States}}
\date{16 January 2007\\ \normalsize \textit{Revised} (v2) \textit{1 May 2007}}

\makeatletter
\renewcommand{\maketitle}{%
  \begin{flushleft}%
    {\LARGE\bfseries\@title\par}%
    \bigskip \bigskip \bigskip \bigskip \bigskip \bigskip
    {\large\@author\par}%
    \medskip
    {\large\@date\par}%
    \bigskip
  \end{flushleft}%
}
\makeatother

\begin{document}
\maketitle

\begin{small}

\noindent A transitional layer matching the asymptotically flat exterior of a charged, massive toroidal body to an interior with spatially cylindrical symmetry is described.  The changes in the geometry, which by themselves would require an energy tensor violating the energy conditions of classical general relativity, are compensated for by an additional strong electric field.  Part of its gradient is consumed, depending on how much the exterior toroid departs from local cylindrical symmetry; what is left over can be used to effect further transitions in the cylindrical interior.  An example is given of such a transition creating an angular deficit, allowing an otherwise infinite cosmic string to be captured within a bounded system. \\

\end{small}

\section{Introduction}

There exist many exact solutions for cylindrically symmetric spacetimes in general relativity.  Some of them model physically mundane sources, such as linear electric currents. \cite{SafkoWitten1972, SafkoWitten1971}  Others, such as cosmic strings, are more exotic. \cite{Hiscock1985, FrolovIsraelUnruh1989, JensenSoleng1992}  Regardless of what values their metric components approach as $\rr \rightarrow \infty$, the infinite extent of their sources in the $\pm z$ direction make them inadequate as approximations to asymptotically flat spacetimes with reasonably bounded sources.  Yet the much greater complexity of coordinate systems adapted to spatially compact bodies which locally approach the symmetry of linelike sources make it difficult to model more realistic analogs.  

In many cases, it seems intuitively obvious that the physical properties of a successfully modelled system with cylindrical symmetry should carry over to the local vicinity of long but finite rods or very thin torii. \cite{Thorne1975} \:  In other cases, where a cylindrical model has some bizarre pathology, those global differences are made into an argument against such generalization. \cite{Visser1996}

In the absence of such solutions, it is still possible to adapt cylindrically symmetric spacetimes as models of bounded bodies by embedding them inside a transitional layer that on the outside is shaped like a finite body, in this case a thin torus.  Because the transitional metric components represent real physical changes in the local geometry, this layer will have a non-vanishing energy tensor.  By themselves, such fields would in places violate the energy conditions of classical general relativity.  However, by including an additional strong field which by itself satisfies the energy conditions but which does not interfere with the transition from one topology to another, the violating contributions of the transitional functions can be compensated for so that $\Tab$ as a whole satisfies these conditions.

\begin{figure}[h]
\begin{center}
\includegraphics[angle=0, height=20pc]{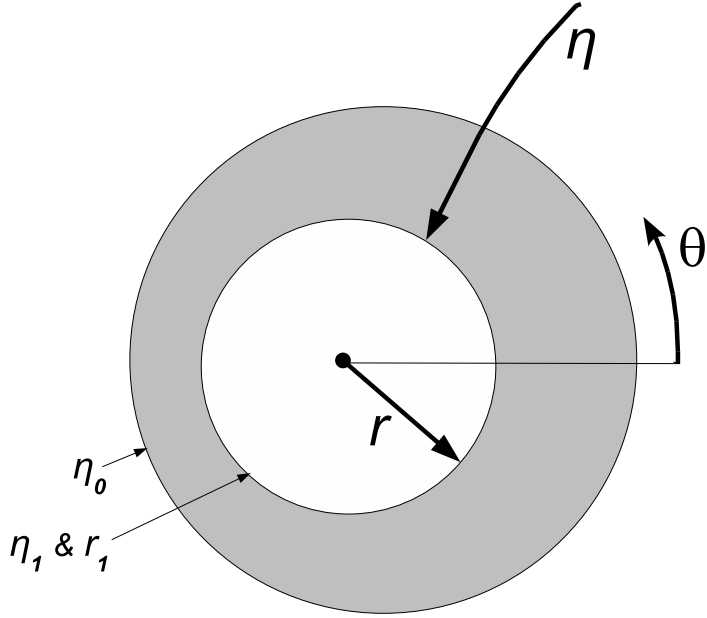}
\caption{Cross-section of transitional ring.  Toroidal exterior, coordinate $0<\eta<\eta_1$ increases inward from zero at spatial infinity; cylindrical interior, coordinate $0<\rr<\rr_1$ increases outward.  External boundary at $\eta_0$, internal boundary at $\eta_1= -\ln (\rr_1/\kappa)$.  Angular coordinate $\theta$ is common to both coordinate patches: $\theta=0$ at the outer equator and $\pm \pi$ at the inner equator of the ring.  The center of the ring is to the left of the figure.}
\label{fig:Toroidal1}
\end{center}
\end{figure}

The exterior spacetime will be that of a thin, nearly uniformly electrically charged massive ring of radius $\kappa$ in the background coordinates \cite{MS}, modelled by the `conformastationary' line element \cite{Majumdar1947, Papapetrou1947, Synge1960}
\be
ds^2 = a^{-2} (dt-w_i \, dx^i)^2 - \frac{\kappa^2 \, a^2}{(\cosh \eta - \cos \theta)^2} (d\eta^2+d\theta^2 + \sinh^2 \! \eta \; d\varphi^2) \label{ds0}
\ee
\bse\be
a(\eta,\theta) = 1 + V(\eta,\theta)
\ee
\be
V(\eta,\theta) = \frac{q}{2 \sqrt{2}} \sqrt{\cosh \eta - \cos \theta} \; P_{-1/2} (\cosh \eta) \label{Vout}
\ee\ese
\noindent where $x^i$ can be any of the spatial coordinates and $P_{-1/2} (x)$ is the Legendre polynomial of order $-1/2$.  Both the ADM mass and the absolute value of the net charge of the ring, as seen by a distant observer, are $\kappa q$.  Among the many advantages of conformastatic spacetimes is their isotropy.  A consequence of this, unlike the more general Weyl spacetimes, is that their physical interpretation is unambiguous: what looks like a ring to a distant observer, still looks like a ring to a local observer.

If $w_i$ is constant it will not appear in any component of $\Tab$, so for convenience it will be set to $0$ in the remainder of this paper.

The signature will be [+ - - -] and geometric units used such that $c_0$, the speed of light in the limit $V \rightarrow 0$, and the gravitational and electric force constants $G$ and $k_e$ are all equal to 1.  $\rt = (x^2+y^2)^{1/2}$ denotes the cylindrical radial coordinate in the interior space.  Where $\rt$ appears as a subscript it denotes a partial derivative, $f_{\rt} = \partial f/\partial \rt$; all other subscripts are labels only.

The definition of standard toroidal coordinates is given in an appendix.

\section{Transitional region}

\subsection{Metric}

Consider the line element
\bea
ds^2_T &=& a^{-2} dt^2 - a^2 \, \kappa^2 \: (N^2 e^{-2\eta} [d\eta^2+d\theta^2] + M^2 d\varphi^2) \label{ds1}
\eea
\noindent with, for now, $a=1$.  $N(\eta,\theta)$ and $M(\eta, \theta)$ are piecewise defined by
\be
N(\eta,\theta) = 
\begin{cases}
  \qquad \qquad \qquad \; \left[\ds\frac{e^{\eta}}{(\cosh \eta - \cos \theta)}\right] & \phantom{\eta_0} 0 < \eta<\eta_0 \vspace{6pt} \\ 
  \; 2 \: \sin^2 \ds\left(\frac{\pi}{2} x\right) + \left[\frac{e^{\eta}}{(\cosh \eta - \cos \theta)}\right] \: \cos^2 \left(\frac{\pi}{2} x\right) & \phantom{0} \eta_0 \leq \eta \leq \eta_1 \vspace{6pt} \\
   \qquad \qquad \qquad \qquad \qquad 2 & \phantom{0} \eta_1 < \eta < \infty
\end{cases} 
\ee
\be
M(\eta,\theta) = 
\begin{cases}
  \qquad \qquad \qquad \;  \left[\ds\frac{\sinh{\eta}}{(\cosh \eta - \cos \theta)}\right] & \phantom{\eta_0} 0 < \eta<\eta_0 \vspace{6pt} \\
  \; \phantom{2 \:} \sin^2 \ds\left(\frac{\pi}{2} x\right) +  \left[\frac{\sinh{\eta}}{(\cosh \eta - \cos \theta)}\right] \: \cos^2 \left(\frac{\pi}{2} x\right) & \phantom{0} \eta_0 \leq \eta \leq \eta_1 \vspace{6pt} \\
   \qquad \qquad \qquad \qquad \qquad 1 & \phantom{0} \eta_1 < \eta < \infty \\
\end{cases} 
\ee

\vspace{12pt}
\noindent where 
\vspace{-24pt}
\bea
x &\equiv& \frac{\eta - \eta_0}{\Delta \eta} \nonumber \\
\Delta \eta &=& \eta_1 - \eta_0  \nonumber 
\eea

For $0<\eta<\eta_0$ (\ref{ds1}) is the Minkowski metric in standard toroidal coordinates (see Appendix).  $\eta$ parameterizes a series of nested torii centered on a ring of coordinate radius $\kappa$, where $\eta \rightarrow \infty$; and $-\pi \leq \theta \leq \pi$ and $0 \leq \varphi \leq 2\pi$ are the angular coordinates of the minor and major orbits respectively.  

For $\eta > \eta_1$ the coordinate transformations 
\bse \bea \label{rz}
\rr &\equiv& 2 \kappa e^{-\eta} \\
z &\equiv& \kappa \varphi \\
\theta &\equiv& \theta
\eea \ese
\noindent show the interior spacetime to be also flat, but locally perfectly cylindrical, with $0\leq z \leq 2 \pi \kappa$ a linear but periodic coordinate, and metric
\bea
ds^2 &=& a^{-2} dt^2 - a^2 (d\rr^2 + dz^2 + \rr^2 \, d\theta^2)
\eea
Examine now the transitional region $\eta_0 \leq \eta \leq \eta_1$, and generalize to the case where $a(x^i)$ is a function of the spatial coordinates other than $\varphi$.  In the frame diagonalizing the pure-electric part of $\Tab$ (not the same as the frame diagonalizing $\Tab$ as a whole)
\be
8\pi a^4 \Tab =  \mathrm{diag} [1,\!\textrm{-}1, \!1, \!1] \, |\nabla a|^2 + \mathrm{diag} [1,0,0,0] \! \left[- 2 a \nabla^2 a -  \!a \nabla a \cdot \negthinspace \nabla \ln M\right] \nonumber 
\ee
\vspace{-18pt}
\be
\qquad \qquad \qquad \qquad \qquad \qquad \qquad \qquad + \; \mathrm{derivatives\;of}\;N,\:M \qquad \qquad \label{Tab} \nonumber
\ee

It can be seen that $\Tab$ can be split into three parts.  The first is a pure electric field, which by itself satisfies the energy conditions.  Its gradient can be adjusted by adjusting the net charge $q$ in (\ref{Vout}) and the outer boundary $\eta_0$.  The third part is due only to the transition functions.  Its magnitude decreases as $\eta_0 \rightarrow \infty$ as the exterior space more closely approaches cylindrical symmetry, for then the relative $\theta$-dependence of $M$ and $N$ decreases and $M \rightarrow 1$, $N \rightarrow 2$.


The middle part contains an $a$-only part and an interaction term.  The latter's relative contribution will also decrease with increasing $\eta_0$.  The Laplacian of $a$ is partly dependent upon the choice of $q$ and $\eta_0$, but it may also be to some extent adjusted independently.  By giving $a$ a small but adjustable negative second derivative, $\Too$ can be made positive and dominant over any other component of $\Tab$.

The two primary constraints on $a$ are that it smoothly ($C^1$) match\footnote{Properly `smooth' requires the metric be $C^{\infty}$ continuous across all boundaries, in other words that it match to all orders of its derivatives.  `Smooth' in this paper refers only to $C^1$ continuity, or continuity in the metric itself and its first derivatives.} with its counterpart on the exterior and that its gradient at $\eta_1$ be positive (or zero, if a flat interior is desired).  The first constraint may be satisfied by
\be
a(\eta,\theta) = 1 + (1-\mu x^2)(C+\Delta \eta \;D\:x) \left(\sin^2 \ds\frac{\pi}{2} x + \left[\frac{{\cosh \eta - \cos \theta}}{{\cosh \eta}}\right]^{1/2} \cos^2 \frac{\pi}{2} x  \right) \nonumber 
\ee
\noindent where the constants $C$ and $D$ are the values at $\eta_0$ of 
\bea
\frac{q}{2\sqrt{2}} \sqrt{\cosh \eta} \; P_{-1/2} (\cosh \eta) \nonumber
\eea
\noindent and its first derivative respectively.  For $\eta_0 \gtrsim 3$ they approach \cite{Hobson1931}
\bse
\bea \label{CD}
C &\approx& (\eta_0 + \ln 4) \: q/2\pi \\
D &\approx& q /2\pi
\eea
\ese
The second constraint is satisfied by restricting 
\be
\mu \leq \mu_{max} = \frac{D \, \Delta \eta}{2 C + 3 D \, \Delta \eta} \approx \frac{\Delta \eta}{2 \eta_0 + 3 \Delta \eta + 2.77} \label{mumax}
\ee 

\subsection{Energy conditions}

For static systems, the pointwise weak, strong, null, and dominant energy conditions are satisfied if for the fully diagonalized energy tensor \cite{HawkingEllis1973}
\bse \label{toroidalec} \bea
\Too \geq 0 \\
\Too \pm T^{\hat{i}\hat{i}} \geq 0 \\
\Too + \Sigma \: T^{\hat{i}\hat{i}} \geq 0
\eea \ese
\noindent Of these seven conditions, the one typically coming closest to violation is $\Too -  T^{\hat{\varphi}\hat{\varphi}}$ at $\eta= \eta_0, \theta = \pi$.  Satisfying this requires that 
\be
\mu \geq \mu_{min} \approx \frac{\phantom{^2}\pi^2}{2} \left(1+\frac{3\pi}{q \eta_0}\right) e^{-\eta_0} \label{mumin}
\ee
\noindent With the upper limit on $\mu$ fixed by (\ref{mumax}), which cannot be greater than $1/3$ in any case, (\ref{mumin}) sets the minimum charge $\kappa q$ a ring of outer boundary $\eta_0$ must have to effect the transition to the cylindrical interior.\footnote{In preparing the table in section \ref{sec:limitations}, $\mu_{min}$ was calculated numerically rather than using the approximate formula (\ref{mumin}).}

Larger values of $\mu$ may be chosen to reduce the gradient of $a$ at the endpoint $\eta_1$ to zero if desired, so as to match the toroidal exterior to a flat and empty interior (section \ref{flat}).  However, the advantage of keeping it as large as possible is that the trick here can be repeated with additional transitional layers to alter other metric functions.  Not only does the cylindrical symmetry of the interior simplify the math needed to do this, it also allows the introduction of features difficult to model in vacuum spacetimes outside bounded sources, such as an angular deficit (section \ref{static string}).

%
%
%
%

\subsection{Optical effects.  A gravitational hall of mirrors}

The interior space is truly cylindrical.  Geodesics propagating parallel to the central axis remain parallel, even though the toroidal exterior has made the interior periodic in the $z$ direction.  A person looking in the $\pm z$ direction would see the back of her own head.  At any given $\rt$, the speed of light in any direction is the same whether one happens to be closer to the `inside' or `outside' edge of the outer ring, since the interior metric includes no dependence on $\theta$.  Yet the spacetime as a whole is asymptotically flat.

%
%
%
%

\section{Limitations}\label{sec:limitations}

At $\eta_1$, relabeled $r_1$ in the new coordinates, $a$ has the form
\bse \bea
a(\eta)|_{\eta_1} &=& 1 + E + F (\eta-\eta_1) \\
E &\equiv& (1-\mu)(C+D\,\Delta \eta) \\
F &\equiv& D (\Delta \eta)^2 - \mu (\Delta \eta) (2 C + 3 D \, \Delta \eta)
\eea \ese
\noindent which after the transformation (\ref{rz}) becomes
\bse \bea
a(\rr)|_{\rr_1} &=& 1 + E - F (r-r_1)/r_1 \label{ar1} \\
&=& 1 + E - F \ln (r/r_1) \label{arlog} \\
E &=& (q/2\pi) (1-\mu) \ln (8\kappa/r_1) \label{E} \\
F &=& (q / 2\pi) \left[(1-\mu) (\Delta \eta)^2 - 2 \mu (\Delta \eta) \ln (8\kappa/r_1) \right]
\eea \ese
If the gradient of $a$ has not been brought to zero at this point, it may be left as (\ref{arlog}), representing the electric field generated by a long wire centered on $\rt=0$ with mass and charge per unit length $F$ (section \ref{chargedwire}).  Alternatively, additional cylindrical shells transitioning to other interior geometries may be placed inside.  The maximum possible distortion of the internal geometry that may be further effected is primarily determined by the relative field gradient at $\rr_1$
\bea
\zeta &\equiv & - (r \, a_r/a)|_{r_1} = F/(1+E) \label{rara}
\eea

Choosing the minimal attenuation of $\nabla a$ from its external value that still satisfies the energy conditions ($\mu = \mu_{min}$) and choosing $\Delta \eta = 1$, the threshold values of $E$ and $\zeta$ for selected $\eta_0$ and $q$, $5 \leq \eta_0 \leq 30$ and $-4 \leq \log_{10} q \leq 4$ are given in the table below.

\arraycolsep=4pt
\ba{r|l|c|c|c|c|c} \nonumber
\eta_0 \phantom{.7}  & E/q & q=10^{-4}             & 10^{-2}                  & 1                           & 10^{2}                  & 10^{4} \\
\hline
30\phantom{.7} & 5.15  & 1.59 \cdot 10^{-5} & 1.51 \cdot 10^{-3} & 2.59 \cdot 10^{-2} & 3.08 \cdot 10^{-2} & 3.09 \cdot 10^{-2} \\
25\phantom{.7} & 4.36 & 1.59 \cdot 10^{-5} & 1.53 \cdot 10^{-3} & 2.97 \cdot 10^{-2} & 3.64 \cdot 10^{-2} & 3.65 \cdot 10^{-2} \\
20\phantom{.7} & 3.56 & 1.59 \cdot 10^{-5} & 1.54 \cdot 10^{-3} & 3.49 \cdot 10^{-2} & 4.46 \cdot 10^{-2} & 4.47 \cdot 10^{-2} \\
15\phantom{.7} & 2.77^* & 1.09 \cdot 10^{-5} & 1.54 \cdot 10^{-3} & 4.22 \cdot 10^{-2} & 5.73 \cdot 10^{-2} & 5.75 \cdot 10^{-2} \\
10\phantom{.7} & 1.97^* &                              & 7.88 \cdot 10^{-4} & 5.30 \cdot 10^{-2} & 7.99 \cdot 10^{-2} & 8.03 \cdot 10^{-2} \\
6.7                     & 1.44^* &                              &                               & 4.89 \cdot 10^{-2} & 9.69 \cdot 10^{-2} & 9.77 \cdot 10^{-2} \\
5\phantom{.7}   & 1.14     &                              &                              &                             & 6.35 \cdot 10^{-2} & 6.51 \cdot 10^{-2}
\ea 

\noindent \small{*For the lowest-$q$ cases in this table, $E/q = 2.74, 1.94, 1.43$ for $\eta_0 = 15, 10, 6.7$ respectively.} \normalsize \\

Since $\mu_{min} \ll 1$, $E$ for each $\eta_0$ is nearly a constant multiple of $q$, as would be expected from (\ref{E}).  As $\eta_0$ decreases and the exterior toroidal surface departs from local cylindrical symmetry to a greater extent, a greater attenuation in $\nabla a$ is required to make the transition and a greater minimal charge $q$ is necessary to keep $\mu_{min} < \mu_{max}$, which is why there are no solutions for the lowest $\eta_0, q$ pairs. 

\begin{figure}[h]
\centering \includegraphics[angle=0, height=19pc]{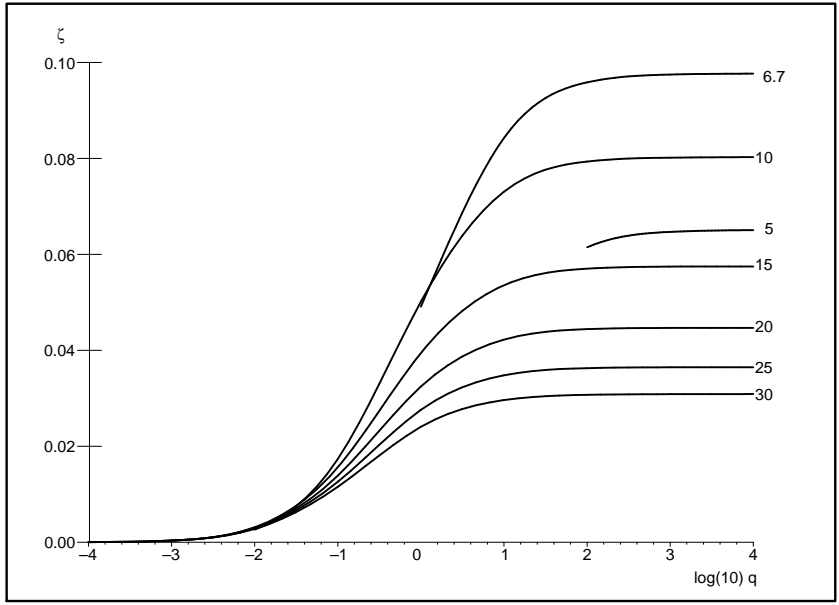}
\caption{Maximum values of $\zeta = -r a_r /a$ at $\rr_1$ as a function of $\log_{10} q$ for selected $\eta_0$.}
\label{fig:rara1}
\end{figure}

As the charge is increased, $\zeta$ approaches asymptotic maxima for each $\eta_0$ (figure \ref{fig:rara1}).  This maximum increases up to about $\eta_0 = 6.7$: the field gradient at $\eta_0$ increases as the torus becomes a more compact body; but beyond this point, the increasing departure from local cylindrical symmetry saps more of the outer field strength in order to effect the transition and $\zeta$, the field strength remaining at $\eta_1$, again declines.  For the choice of transition functions used here the maximum possible $\zeta$ is less than $0.10$, although minor tweaking of the transition functions can get it above that threshold.\footnote{For instance, globally replacing $\sin^2 (\frac{\pi}{2}x) \to \sin^4 (\frac{\pi}{2}x)$, $\cos^2 (\frac{\pi}{2}x) \to 1-\sin^4 (\frac{\pi}{2}x)$, which reduces the curvature of $N$ and $M$ at $\eta_0$ where (\ref{toroidalec}) come closest to violation.}

%
%
%
%

\section{Examples}

%
%
%
%

\subsection{Charged wire} \label{chargedwire}

Leaving $a(\rt)$ in the logarithmic form (\ref{arlog}) makes the interior spacetime vacuum except for a cylindrically symmetric electric field, representing (after matching to a suitable second interior solution at some $\rB < \rr_1$) the exterior to a long wire centered at $\rt=0$ with a linear mass and charge density $Q=F$.  As noted above, the periodicity of $z$ means that the wire is not really infinitely long, although the metric, energy tensor, and most local measurements will be identical with those of a wire that is.

%
%
%
%

\subsection{Flat space} \label{flat}

\begin{figure}[t]
\begin{center}
\includegraphics[angle=0, width=\textwidth]{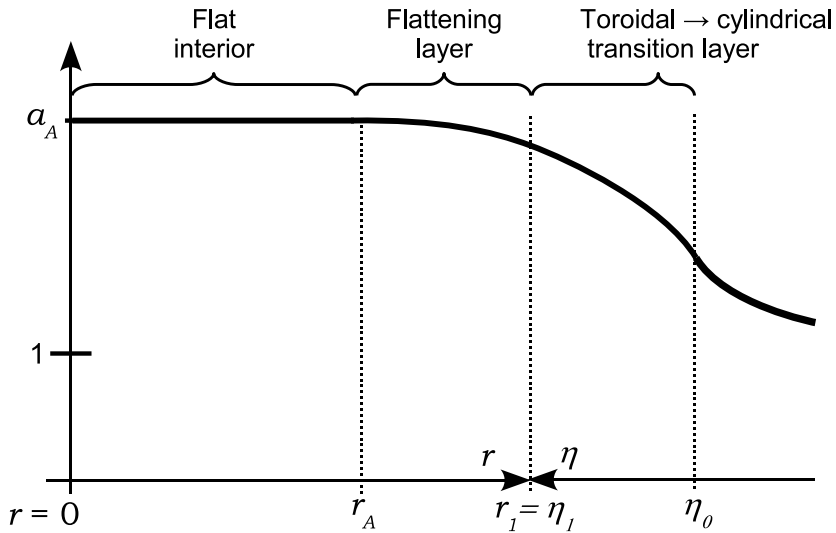}
\caption[$a(\rr)$, $a(\eta)$ for the example of section \ref{flat}.]{$a(\rr)$, $a(\eta)$ for the example of section \ref{flat}.  A second transitional shell covering $\rA < \rr < \rr_1$ is placed within the original toroidal-to-cylindrical transitional shell, eliminating the remaining gradient of $a(\rr)$ to create a flat space inside.}
\label{fig:aflat}
\end{center}
\end{figure}

Truncating $\nabla a \rightarrow 0$ at some $\rA < \rt_1$ will match the spacetime to a flat interior (figure \ref{fig:aflat}).  A suitable function matching to $a(\rr_1)$ (eq. (\ref{ar1})) and $a(r)=a_{\scriptscriptstyle{A}}$ = constant at $\rA$ is
\bea
a(r) &=& a_{\scriptscriptstyle{A}} (1-\nu y^2) \\
\nu &=&\frac{\zeta \Delta \rr}{2 \rr_1 - \zeta \Delta \rr} \\
y &\equiv& (\rr - \rA)/(\Delta \rr)  \nonumber \\
\Delta \rr &=& \rr_1 - \rA  \nonumber 
\eea

The interior space is flat and isotropic, though still periodic in the $z$-direction.  Spatial distances are greater and time runs slower relative to the outside by a factor of $a_{\scriptscriptstyle{A}} = (1-\nu)^{-1} (1+E)$.  The local speed of light in all directions is $(1/a_{\scriptscriptstyle{A}})^{2}$.

%
%
%
%

\subsection{A captive string} \label{static string}

\begin{figure}[t]
\begin{center}
\includegraphics[angle=0, width=\textwidth]{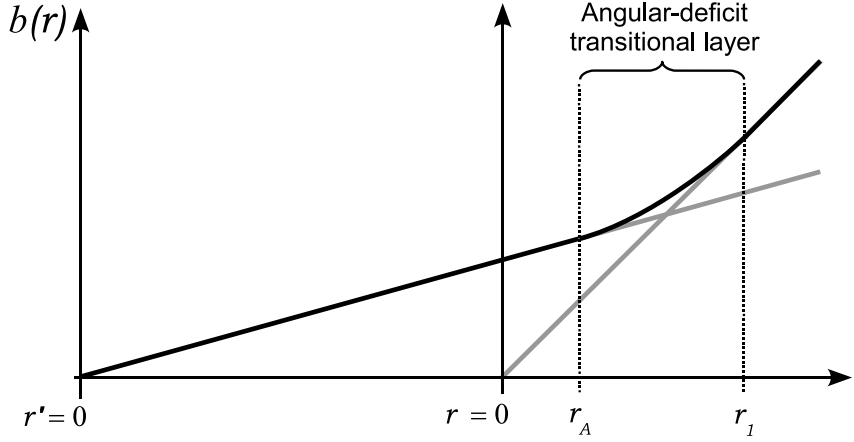}
\caption[$b(\rr)=\sqrt{-g_{\theta \theta}}$ for the example of section \ref{static string}.]{$b(\rr)=\sqrt{-g_{\theta \theta}}$ for the example of section \ref{static string}.  The slope of $\sqrt{-g_{\theta \theta}}$, $d b(\rr)/d\rr$, is attenuated from 1 to a lesser value over $\rA < \rr < \rr_1$ to create an angular deficit within the interior space.  Since $\rr' \equiv \rr+\rr_s$, the true origin $\rr'=0$ or $\rr=-\rr_s$, where the cosmic string is located, is shifted inward from $\rr=0$.  }
\label{fig:stringtransition}
\end{center}
\end{figure}

\noindent By transitioning to an interior with an angular deficit, a cosmic string with its exact vacuum solution may in effect be captured within a spatially bounded system (figure \ref{fig:stringtransition}).  Write the line element as
\bea
ds^2 &=& a^{-2} dt^2 - a^2 (dr^2+dz^2+b(r)^2 d\theta^2)
\eea
\noindent with $a$, $\Delta r$, and $y$ defined as in section (\ref{flat}) and $b(r)$ defined as\footnote{Properly since $\rr_1$ marks the outer boundary of the cylindrical interior space, the coordinate $\rr$ should not be continued beyond it as in figure \ref{fig:stringtransition} and equation (\ref{brstring}).  This coordinate abuse is done for clarity of exposition.}
\be \label{brstring}
b(r) = 
\begin{cases}
 \quad \rr & \quad \rr > \rr_1 \\
 \quad \rr + \sigma (1-y)^2 & \quad \rA \leq \rr \leq \rr_1 \\
 \quad B (\rr+\rr_s) & \quad \rr < \rA \\
\end{cases} 
\ee
\bse \bea
\sigma &=& \left(\frac{1-B}{2}\right) \Delta \rr \label{N} \\
\rr_s &=& \left(\frac{1-B}{B}\right) \frac{\rA + \rr_1}{2}
\eea \ese

For $B \equiv 1-4\lambda <1$ and the coordinate substitution $\rr' \equiv \rr + \rr_s$, the spacetime inside $\rr<\rA$ is the exterior to a straight, static cosmic string with linear mass density $\lambda$, together with a radial electric field.

The general energy tensor has again an electric part $\Tab_{\scriptscriptstyle{E}} = (8\pi a^4)^{-1} |\nabla a|^2$ $\times$ diag$[1,-1,1,1]$ plus
\bse \bea
\Too - \Too_{\scriptscriptstyle{E}} &=& (8\pi a^3 b)^{-1} (-2 b a_{\rr \rr} - a b_{\rr \rr} - 2 b_{\rr} a_{\rr}) \\
T^{\hat{z}\hat{z}} - T^{\hat{z}\hat{z}}_{\scriptscriptstyle{E}} &=& (8\pi a^3 b)^{-1} a b_{\rr \rr} 
\eea \ese
\noindent With $a$, $b$, $\sigma$, $b_r$, $b_{rr} > 0$ and $a_r, \: a_{rr} < 0$, the energy conditions will be satisfied if $\Too - T^{\hat{z}\hat{z}} \geq 0$, or
\bea
-a_{rr} \geq b^{-1} (a b_{rr} + b_r a_r) \label{staticstringEC}
\eea
Again, the energy deficit will be met by attenuating $\nabla a$ from its value at $\rr_1$.  For simplicity, $\nabla a$ will be reduced to $0$ at $\rA$ as in section \ref{flat}, eliminating the electric field and leaving only the vacuum exterior to a string located at $r'=0$.  In other words, the entire amount of $\zeta$ will be spent on creating the largest possible angular deficit.  Then (\ref{staticstringEC}) will be satisfied so long as $\sigma = \alpha \, \rA \nu$, $\alpha \leq 1$.  Equating this $\sigma$ to (\ref{N}), and choosing $\alpha=1$ and $\Delta \rr \ll \rr_1$ so that $\rA \approx \rr_1$, gives for the maximum possible angular deficit
\bse \bea
\Delta \theta_{max}&=& 2 \pi (1-B_{max}) \\
&=& 2 \pi \zeta
\eea \ese

\noindent allowing at best a relative reduction of 10\% in $\sqrt{g_{\theta \theta}}$ under the constraints of this model.

Returning to the original form of the exterior metric (\ref{ds0}), it can be seen that a spinning string, so long as it is rotating around its longitudinal axis, can be modelled as well: the introduction of a constant-coefficient term $w_{\theta} \, d\theta$ has no effect on the energy tensor anywhere, including the transition layer and the exterior toroidal spacetime.  

A rotating loop of string could also be modelled with the introduction of a constant-coefficient $w_z \, dz$ metric term in the internal coordinates, with a corresponding constant-coefficient $w_{\varphi} \, d\varphi$ in the exterior metric.  But then the outer spacetime would have a conical singularity along the axis of the ring and so would no longer be asymptotically flat.  

In general, the upper limit on the relative field gradient $\zeta$ is the chief restriction on what cylindrical interiors may be embedded within these mediating rings.  A transitional layer designed to contain a van Stockum cylinder \cite{vanStockum1937}, for instance, would require using up part of $\zeta$ on bringing $w(r) = -g_{t \theta} / g_{tt}$ to a non-constant value and another part on creating an angular deficit.  This could be done for a causally well behaved interior solution; but the maximum possible $\zeta \lesssim 0.10$ is too weak to bring these parameters to values that would introduce closed timelike curves.

\section*{Acknowledgements}

I am grateful to A Kheyfets and S V Krasnikov for advice and criticism related to the content of this paper.  Calculations were largely done with GRTensor II \cite{grtensor} on Maple 5 and 10.  This work was partially supported by the National Security Education Program.

%
%
%
%

\begin{appendix}

\section*{Appendix.  Standard toroidal coordinates}

\begin{figure}[t]
\begin{center}
\includegraphics[angle=0, width=\textwidth]{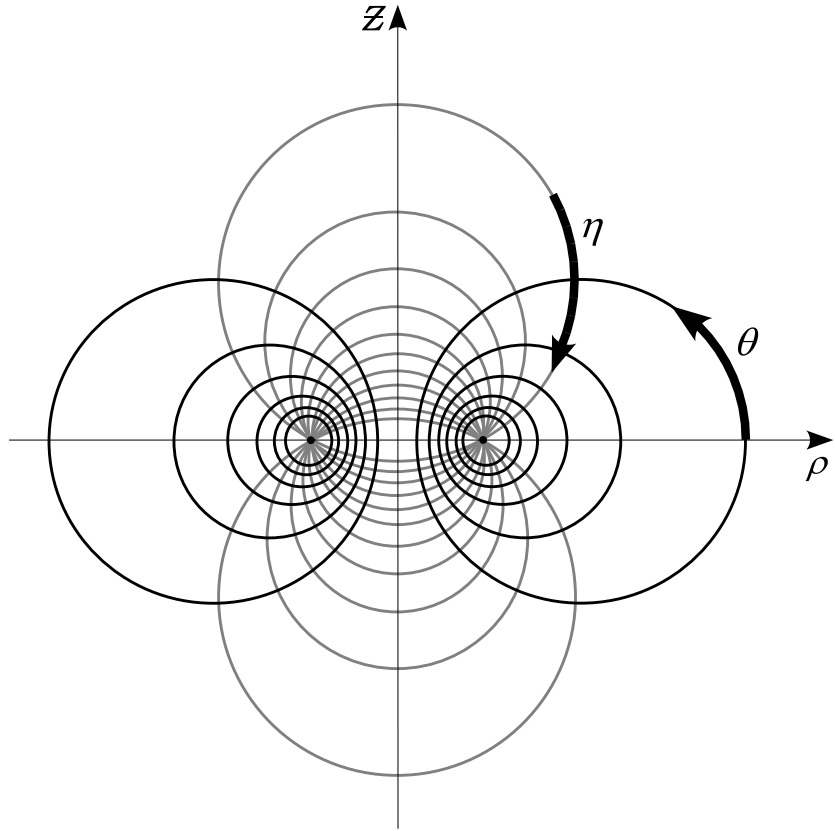}
\caption{Standard toroidal coordinates, constant-$\varphi$ cross-section.  Central ring located at $Z\hspace{-7pt}-\!=\!0, \, \rho\!=\!\kappa$ (black dots).  Radial-type coordinate $\eta$ (constant-$\eta$ surfaces are black lines) increases from 0 on the vertical axis and at spatial infinity to $\infty$ on the core of the ring; angular coordinate $\theta$ (grey lines) increases from 0 on the $Z\hspace{-7pt}-=0$ plane outside the ring, to $\pm \pi$ on the central disk spanning it.  Adapted from \cite{Wikimedia}.}
\label{fig:toroidalcoordinates}
\end{center}
\end{figure}

The definitions of standard toroidal coordinates are given here, rather than in the main body of the text, to avoid confusion between the cylindrical coordinate system $(\rho, \ZZ, \varphi)$ in the external space by which they are most easily defined and the cylindrical coordinate system $(r, z, \theta)$ used throughout this paper for the cylindrical space within the toroidal shell.

With reference to a cylindrical coordinate system in the exterior spacetime (figure \ref{fig:toroidalcoordinates}), toroidal coordinates can be implicitly defined by
\vspace{-6pt}
\bea
\rho &=&  \frac{\kappa \:\sinh \,\eta}{\cosh \eta - \cos \theta} \\
\ZZ &=& \frac{\kappa \: \sin \,\theta}{\cosh \eta - \cos \theta}
\eea

\noindent where $\kappa$ is the radius of the source ring.  They have a flat three-metric
\bea
ds^2 &=& \kappa^2 (\cosh \eta - \cos \theta)^{-2} \left[ (d\eta^2+d\theta^2) +  \sinh^2 \eta \:  d\varphi^2 \right]
\eea

\noindent Axisymmetric solutions to Laplace's equation suitable for the exterior to a source have the form
\ba{lll}
f_n (\eta,\theta) &=& (\cosh \eta - \cos \theta)^{1/2} P_{n-1/2} (\cosh \eta) \: (A \cos n \theta + B \sin n \theta)
\ea

\noindent They are built from Legendre polynomials of half-integer order, sometimes called `toroidal harmonics'.  These do not have a closed form.  For $n\geq0$ they become proportional to $e^{-(n+1/2)\eta}$ for $\eta \to \infty$. \cite{Hobson1931}

Note that the solutions to Laplace's equation are no longer simply separable (they belong to the `R-separable' class of solutions \cite{MS}).  Even the zero-order solution will have some relative dependence on the angular coordinate $\theta$.  But like the $(\cosh \eta - \cos \theta)$ term in the metric itself, this dependence will rapidly decrease as $\eta$ gets large.

For further discussion of toroidal coordinates, toroidal harmonics, and modelling the exterior fields of ring sources, see \cite{MS, Hobson1931}.

\end{appendix}

\end{document}